# Inferring three-body interactions in cell migration dynamics

Agathe Jouneau,[1] Tom Brandstätter,[2,3] Bram Hoogland,[2] Joachim O. Rädler,[1] and Chase P. Broedersz[2]

[1]*Faculty of Physics and Center for NanoScience, Ludwig-Maximilians-Universität München, 80539 Munich, Germany*
[2]*Department of Physics and Astronomy, Vrije Universiteit Amsterdam, 1081 HV Amsterdam, The Netherlands*
[3]*Arnold-Sommerfeld-Center for Theoretical Physics, Ludwig-Maximilians-Universität München, 80333 Munich, Germany*

In active matter and living matter, such as clusters of migrating cells, collective dynamics emerges from the underlying interactions. A common assumption of theoretical descriptions of collective cell migration is that these interactions are pairwise additive. It remains unclear, however, if the dynamics of groups of cells is solely determined by pairwise interactions, or if higher-order interaction terms come into play. To investigate this question, we use time-lapse microscopy to record the dynamics of three cells interacting together in a linear three-site geometry. We collect a large number of cellular trajectories and develop an inference scheme to infer both pairwise and potential three-body cell-cell interactions. Our results reveal evidence of three-body interactions in one of the two cell lines tested. However, these three-body interactions only introduce minor corrections to the overall dynamics. Our work provides a methodology to infer the existence of three-body interactions from trajectory data, and supports the commonly assumed pairwise nature of cell-cell interactions.

*Introduction*—Many animal cells have the remarkable ability to migrate in a coordinated and collective manner. Collective cell migration is central in biological processes, including embryonic development, wound healing, and cancer metastasis [1–3]. Multicellular systems are also a prominent example of organized active matter [4–6], and there has been an effort to build physical models that capture their dynamics [7–9]. One popular approach uses 'active particle models', where each cell is treated as a point particle subject to active forces accounting for its motility and interactions with neighboring cells [7,9]. Despite their apparent simplicity, active particle models have been able to provide a quantitative description of migrating cells in various contexts [10–17]. However, it remains challenging to determine the exact nature of cell-cell interactions in such models from the underlying biomolecular machinery [8,18,19]. These interactions are therefore often constructed on phenomenological grounds with limiting assumptions.

A central assumption in most models is that cell-cell interactions are pairwise additive [9–11,14–16]. However, since cellular interactions integrate physical forces and biochemical signaling mechanisms [3,18,19], it is unclear whether higher-order interactions, involving three or more cells, may be at play. Higher-order interactions could affect the emergent properties of multicellular systems, as in other biological and soft matter systems [20–27]. Clarifying their presence and importance is therefore crucial for understanding the dynamics of multicellular systems [11,15,17,19,28].

Detecting higher-order interactions from noisy experimental trajectories is challenging [25]. For colloidal suspensions at thermal equilibrium, image-based data has successfully been used to derive the three-body interaction between particles [29]. For cell migration, inference methods have proven to be a powerful tool for determining cell-cell interactions from measured cellular trajectories [19]. A popular approach to infer cell-cell interactions is to confine the cells in a microenvironment [12,14,17,19,30–32]. In particular, micropatterns can be employed as 'cell colliders' to systematically infer pairwise cell-cell interactions from experimental data [12].

Here, we develop a high-throughput method to study the complex dynamics of groups of three cells on confining micropatterns. We generalize the framework of Underdamped Langevin Inference (ULI) [33] to infer three-body cell-cell interactions directly from the measured cellular trajectories. We examine two breast-derived cell lines: a noncancerous epithelial cell line (MCF10A) and a cancerous mesenchymal cell line (MDA-MB-231). Interestingly, we find that the dynamics of the first is fully captured by pairwise interactions, while the second exhibits distinct three-body interactions. Importantly however, multicellular dynamics remains largely dominated by pairwise interactions, even for the cancerous cells. Our approach can be applied to other systems and geometries, where three-body interactions may play a more dominant role.

*Dynamics of three interacting cells*—To investigate the presence of three-body interactions in migrating cells, we record the trajectories of three cells interacting within a confined microenvironment. This environment consists of three square islands connected by two thin bridges (Fig. 1(a)), extending a geometry previously used to study two-cell interactions [12]. We produce large arrays of confining micropatterns coated with fibronectin, a protein that promotes cellular adhesion. The outer area is passivated with PLL-PEG, which prevents cells from adhering.

We select patterns in which three cells adhere and record their dynamics using time-lapse microscopy for up to 48 h. During the experiment, we observe that the cells are in contact through their extended protrusions and regularly move between the square sites (Fig. 1(b)). This protocol thus allows us to study dynamics of a minimal multicellular system in which three-body interactions could manifest.

We compare the epithelial cell line MCF10A, and the cancerous mesenchymal cell line MDA-MB-231, both derived from human breast tissue. We observe a qualitative difference in the collective behavior of the two cell lines: while MCF10A cells favor configurations where cells cluster together, often sharing the same site, MDA-MB-231 cells prefer to occupy separate sites. The MDA-MB-231 cells also exhibit faster dynamics than the MCF10A cells, moving more frequently from site to site and reaching higher velocities (see [34] for movies).

To quantify this multicellular behavior, we determine one-dimensional cellular trajectories. For each cell, we use the position of the nucleus center to determine its position $x(t)$ along the pattern's long axis (Fig. 1(c)). For each cell line, we collect over 1500 hours of trajectories of three cells interacting under standardized conditions.

*Stochastic equation of motion*—Previous works showed that the dynamics of migrating cells can be captured by an underdamped stochastic equation of motion [10,12,35–37]. In this framework, the acceleration of a cell $i$ is given by

$$\frac{dv_i}{dt} = F_i(\mathbf{x}, \mathbf{v}) + \sigma \eta_i(t) \quad (1)$$

where $\mathbf{x} = \{x_1, x_2, x_3\}$ and $\mathbf{v} = \{v_1, v_2, v_3\}$ are the positions and velocities of the three cells along the $x$-axis. $F_i$ represents a deterministic effective force, and $\eta_i$ a Gaussian white noise ($\langle \eta_i(t) \rangle = 0$, $\langle \eta_i(t)\eta_j(t') \rangle = \delta(t-t')\delta_{ij}$) with amplitude $\sigma$. For single cells migrating within a confinement, $F_i = F_{\text{ext}}$ describes the dynamics arising from the interaction with the confinement [35,38,39]. For two interacting cells, a cell-cell interaction term $F_{2B}$ is added [12].

In a multicellular system with three – or more – interacting cells, a natural extension would be to sum the cell-cell interaction term $F_{2B}$ over each pair of cells, as assumed in many models [9–11,14–16]. However, there is no *a priori* reason for such a living system to behave in a pairwise additive manner. Indeed, cell-cell interactions rely on a complex integration of physical forces and signaling mechanisms [3,18]. Therefore, higher-order interactions such as three-body interactions, may be necessary to fully capture their dynamics. We define a three-body interaction as an interaction that depends on the positions and velocities of all three cells and cannot be decomposed into a sum of pairwise interactions.

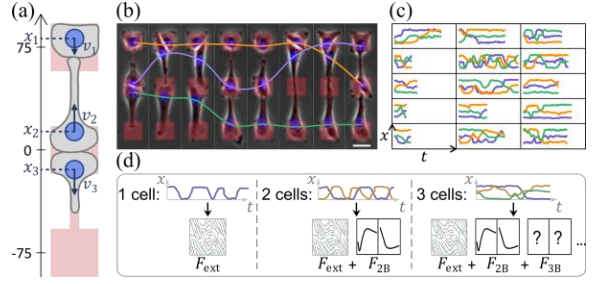

FIG. 1. Stochastic dynamics of three interacting cells. (a) Sketch of three cells confined on a three-site micropattern. (axis unit: μm). (b) Time-lapse microscopy images for MDA-MB-231 cells. Micropatterns are labelled in red, and cell nuclei in blue. Time interval between two images: 40 minutes. Scale bar: 30 μm. (c) Sample set of cellular trajectories as a function of time (0 < t < 20 h). (d) Schematic of the inference framework applied to systems with one [32], two [12] and three cells migrating on a confining micropattern. Underdamped Langevin Inference (ULI) is used to infer cell interaction with the micropattern ($F_{\text{ext}}$), two-body interactions ($F_{2B}$) and potential three-body interactions ($F_{3B}$).

Then, the global effective force becomes a superposition of three terms: the first describing the cells' intrinsic motility within the environment; the second describing the cells' pairwise interactions; and a three-body interaction term:

$$F_i(\mathbf{x}, \mathbf{v}) = F_{\text{ext},i} + \sum_{j \neq i} F_{2B,ij} + \sum_{i \neq j, i \neq k, j < k} F_{3B,ijk} \quad (2)$$

*Inferring three-body interactions*—We use Underdamped Langevin Inference (ULI), which enables a rigorous and robust inference of deterministic effective forces and noise amplitude from experimental trajectories [33]. It has previously been used to infer $F_{\text{ext}}$ in the case of single-cell migration [33], and to simultaneously infer both $F_{\text{ext}}$ and $F_{2B}$ from data of two interacting cells [12]. We generalize this approach and adapt ULI to simultaneously infer $F_{\text{ext}}$, $F_{2B}$ and $F_{3B}$ directly from our experimental data (Fig. 1(d)).

ULI employs projections of the accelerations onto a set of basis functions and corrects for systematic errors arising from finite sampling [33]. Given a basis expansion for $F_{\text{ext}}$, $F_{2B}$ and $F_{3B}$, ULI infers the associated expansion coefficients such that the resulting function best explains the data. An intrinsic trade-off of the ULI method is that the inference gets less reliable when the set of basis functions gets larger. This makes the choice of the basis both subtle and critical [19]. For $F_{\text{ext}}$ and $F_{2B}$, we use the basis functions that were previously established [12].

For $F_{3B}$, we reduce the dimensionality of the inference by considering the symmetries of the system. Specifically, we restrict ourselves to functions $F_{3B}(\Delta x_{ij}, \Delta x_{ik})$ where $\Delta x_{ij} = x_i - x_j$, which satisfy the following conditions: $F_{3B}(-\Delta x_{ij}, -\Delta x_{ik}) = -F_{3B}(\Delta x_{ij}, \Delta x_{ik})$ (symmetry under reversal of the $x$-axis) and $F_{3B}(\Delta x_{ik}, \Delta x_{ij}) = F_{3B}(\Delta x_{ij}, \Delta x_{ik})$ (identical cells). We further reduce dimensionality by using simplifying ansatzes that consider three-cell interactions in various distinct configurations of cells. This reduces the inference to that of a one-variable function, as detailed in the Supplemental Material (SM) [34].

We use bootstrapping to estimate $3\sigma$-error bars for the terms inferred by ULI [34,40,41]. To test the accuracy of the three-body interaction inference, we simulate trajectories with a given three-body interaction, perform inference with bootstrapping, and verify that the original input three-body interaction is encompassed by the $3\sigma$-error bar (Fig. S1(a)). We repeat this process with independent, randomly chosen three-body interactions and find that the input interaction lies within the bootstrap error bar in at least 95% of cases. This demonstrates that we can reliably re-infer a three-body interaction, despite large error bars (Fig. S1(b) and Fig. S2) [34]. Based on these results, we consider a three-body interaction to be significant if its $3\sigma$-error bar does not encompass zero.

*Two-body interactions are conserved in a three-cell system*—Using our experimental data of three interacting cells, we infer $F_i(\mathbf{x}, \mathbf{v})$ (Eq. 2). We first examine two-body cell-cell interaction $F_{2B}$. For two cells interacting on a two-island pattern, it was shown that the cell-cell interaction term can be written as a sum of two contributions: $F_{2B,ij} = f(|\Delta x_{ij}|)\Delta x_{ij} + \gamma(|\Delta x_{ij}|)\Delta v_{ij}$ where $\Delta x_{ij} = x_i - x_j$ and $\Delta v_{ij} = v_i - v_j$. The function $f$ can be interpreted as an effective attraction, and $\gamma$ as an effective cell-cell friction. Both are found to vary between different cell lines [12,17].

Notably, we find that for both cell lines, the two-body interactions $f$ and $\gamma$ as inferred from our three-cell systems, are almost identical to the ones inferred for two cells interacting in a two-island pattern [17] (Fig. 2 (a to d)). More specifically, we infer short-range repulsion and friction for the MCF10A cells, and short-range attraction, long-range repulsion and anti-friction for the MDA-MB-231 cells. There are some quantitative differences with prior work on two cells [12], which we attribute to the difference in the patterning technique (Fig. S3) [34]. The term $F_{\text{ext}}$ and the noise amplitude $\sigma$ are also similar to the ones inferred for two cells on the two-island pattern (Fig. S4) [34]. Finally, we note that the two-body interactions inferred from three-cell systems remain nearly unchanged if a three-body interaction is included in the inference (Fig. 2 (a to d)).

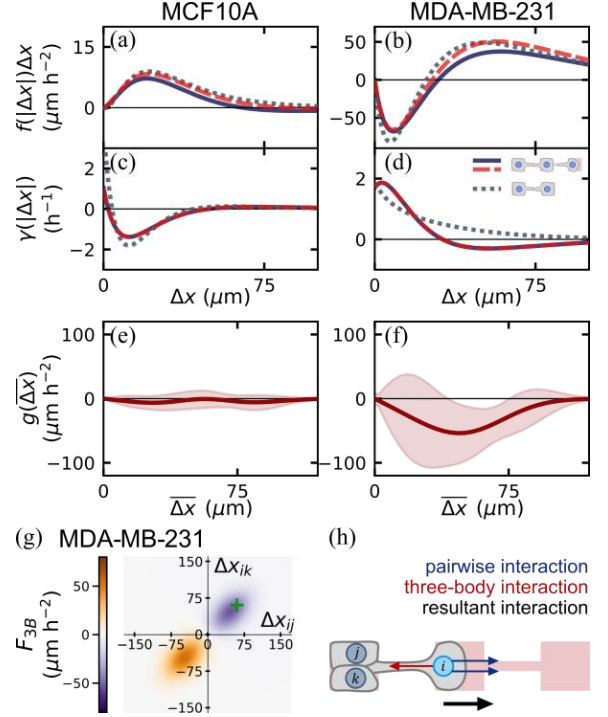

FIG. 2. Inferred pairwise and three-body interactions. (a and b) Pairwise cohesive and (c and d) pairwise frictional interaction terms, inferred for MCF10A and MDA-MB-231 cells. Solid blue lines are the interactions inferred from a purely pairwise model and dashed red lines when the three-body interaction is included. Gray dotted lines are the interactions inferred for two cells interacting in a two-island confinement [17]. (e and f) Inferred three-body interaction for MCF10A and MDA-MB-231 cells, with bootstrap $3\sigma$-error bar (light red area). (g) 2D-representation of the three-body interaction inferred for MDA-MB-231 cells. (h) Schematic of the configuration indicated by the green cross in (g).

*A three-body interaction is detected for MDA-MB-231 cells*—Next, we consider the inferred three-body interactions. For simplicity, we first present the results obtained with one of the simplest ansatzes for three-body interactions, for which the inferred interaction is also the most significant. This ansatz describes a three-body interaction that arises when two cells are in close proximity and together affect the third cell. In this configuration, the repulsive or attractive forces exerted by each cell may no longer be pairwise additive. The deviation from pairwise additivity can be captured by a three-body interaction, of the form $F_{3B,ijk} = g\left(\frac{\Delta x_{ij}+\Delta x_{ik}}{2}\right)e^{-\frac{\Delta x_{jk}^2}{2w^2}}$. When cells $j$ and $k$ are close to each other (i.e. $\Delta x_{jk}$ is small), the function $g$ describes the positive or negative amplitude of the deviation as a function of the average distance $\overline{\Delta x}$ between cells $j$ and $k$, and cell $i$.

We find two qualitatively different results for the two cell lines: For the MCF10A cell line, we find no significant three-body interactions (Fig. 2(e)). In contrast, for the MDA-MB-231 cell line, we reliably infer a non-zero three-body interaction (Fig. 2(f)). This three-body interaction is attractive in nature, and we interpret it as a correction to the additivity of two-body cell-cell repulsion, such that the global repulsion exerted by two colocalized cells is weaker than the sum of the pairwise repulsions (Fig. 2(g)).

Our inference results suggest that the dynamics of MCF10A cells is driven by two-body interactions, whereas a three-body interaction is present for MDA-MB-231 cells. This raises the question of how this three-body interaction impacts their collective dynamics.

*Cellular dynamics is dominated by two-body interactions*—To check the predictive power of our inferred model and assess the role of three-body interactions in MDA-MB-231 cells, we simulate Eq. 1 with the inferred terms and compare experimental and simulated statistics. Since the inference uses only projections of cell accelerations, the predictive power of the model must be tested using different, longer-term statistics. For the MCF10A cells, based on our inference results, we simulate a simple pairwise interaction model. For the MDA-MB-231 cells, we compare a model including the inferred three-body interaction to a purely pairwise model.

We compare the statistical properties of simulated and experimental trajectories, such as the steady-state distributions of positions, velocities and dwell times (Fig. 3 (a to f), Fig. S5 and Fig. S6) [34]. For the MCF10A cell line, all experimental statistics are well captured by a model including only pairwise interactions. For the MDA-MB-231 cell line, we observe little difference between the model that includes the three-body interaction and the pairwise model, and both models provide a good match to the statistics.

To better assess the influence of the three-body interactions, we consider two additional statistics that we expect to be particularly sensitive to cell-cell interactions. The first one is the distribution of relative positions, also called triplet correlation function, which has shown to be sensitive to three-body interactions for colloidal suspensions [29]. It is defined as the probability of finding three particles at distances $r_{12}$, $r_{13}$ and $r_{23}$ from each other. In one dimension, it reduces to two relative positions: $\Delta x_{12}$ and $\Delta x_{13}$ ($\Delta x_{23} = \Delta x_{13} - \Delta x_{12}$). The second statistic is the transition rate between different coarse-grained cellular configurations (Fig. S8) [34], and quantifies cell rearrangements.

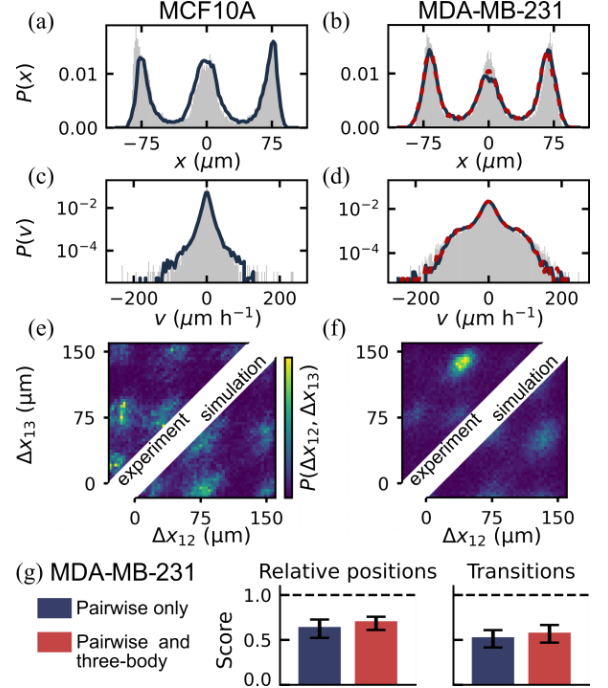

FIG. 3. Comparison of simulated and experimental data: (a and b) Cell positions and (c and d) cell velocities (semi-log plot) for MCF10A and MDA-MB-231 cells. The gray areas show the experimental distribution, and continuous blue and red dashed lines the distributions from the pairwise interaction model and from the model including the three-body interaction, respectively. (e and f) Joint distribution of the relative cell positions. The upper (lower) triangle shows the distribution of experimental (simulated) data, for MCF10A cells (pairwise interaction model) and MDA-MB-231 cells (model with three-body interaction). (g) Scores of how well the relative positions and transitions between cellular configurations are captured by the models, for MDA-MB-231 cells. Higher scores indicate better agreement between simulation and experiment.

We observe that both these statistics are reasonably well captured for both cell lines (Fig. 3, Fig. S7 and Fig. S8) [34]. For the MDA-MB-231 cell line, we evaluate the improvement of including a three-body interaction in the model by computing scores for how well cell relative positions and cell transitions are captured (Fig. 3(g)). These scores are derived from the distances between the associated distributions, with a score of one indicating perfect alignment with the experiment. We do not observe a statistically significant improvement, and therefore conclude that the three-body interaction detected in MDA-MB-231cells constitutes a minor correction to the dynamics.

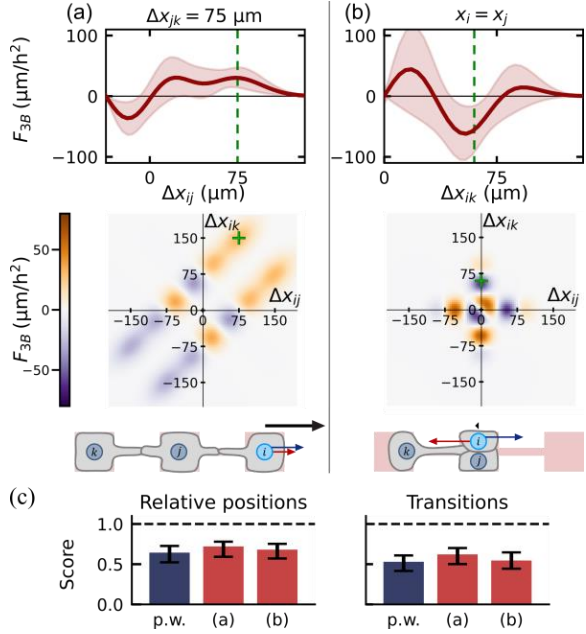

FIG. 4. Inference results for different implementations of three-body interaction, for MDA-MB-231 cells. (a and b) For two additional implementations, the first line shows the inferred function $F_{3B}$ with bootstrap $3\sigma$-error bar, the second line shows a 2D-representation of $F_{3B}(\Delta x_{ij}, \Delta x_{ik})$, and the sketches illustrate how the three-body interaction (red arrow) adds to the pairwise interaction (blue arrow) and contributes to the resultant interaction (black arrow) acting on cell $i$, in the configuration indicated by the green dashed line and cross. (c) Scores of relative positions and transitions between cellular configurations of simulated data are compared to experiments, for a pairwise interaction model (p.w.) and for models including three-body interactions depicted in (a) and (b), respectively.

*Other three-body interactions*—We explore several other possible ansatzes for three-body interactions. For the MCF10A cells, we find no significant three-body interactions for any of the considered ansatzes (Fig. S9) [34]. However, for the MDA-MB-231 cells, we infer other significant three-body interactions, which are shown in Fig. 4. When considering a configuration in which cells $j$ and $k$ are one island apart, we infer a three-body repulsion when cell $i$ is located to the left (Fig. 4(b)). Additionally, when considering a configuration in which cells $i$ and $j$ are colocalized, we infer a three-body attraction with a maximum when cell $k$ is located 55 μm to the left (Fig. 4(c)). Other three-body interactions inferred using different ansatzes are equivalent to the ones cited (Fig. S9) [34]. As before, including these three-body interactions does not significantly improve the scores measuring how well the relative positions and transitions between cell configurations are captured (Fig. 4(c)). The same applies to all the other ansatzes we tried, including a direct inference of a general two-variable function of cell distances and three-body interactions that depend linearly on the cellular velocities (Fig. S10) [34]. We conclude that, for both cell lines, the dynamics is dominated by two-body interactions, and that the three-body interactions detected in the MDA-MB-231 cell line are minor corrections to the dynamics.

*Discussion*—We investigated the existence and significance of three-body interactions in a biological system. For this purpose, we developed a method that allows us to infer the presence of three-body interactions between cells based on their stochastic trajectories. We applied this method to two different cell lines and detected significant three-body interactions only for the cancerous mesenchymal cell line (MDA-MB-231). These inferred three-body interactions introduce a correction to the pairwise additivity of two-body interactions and can be attractive or repulsive depending on the configurations. Nevertheless, we found that the multicellular dynamics is well captured by a model that only considers pairwise interactions, indicating that two-body interactions dominate the behavior.

While we cannot exclude the possibility that higher-order cellular interactions play a more dominant role in two- or three-dimensional geometries involving more interfacial contact between cells, our results corroborate the common use of the pairwise-interaction assumption in the field of cell migration. These findings are reminiscent of other physical systems in which three-body interactions appear to be higher-order corrections to dynamics, for example in atoms interacting via van der Waals forces [42,43], or in colloidal suspensions [29]. Our method could be adapted to explore three-body interactions in a broader range of active matter systems with stochastic dynamics.

*Acknowledgments*—We thank Emily Brieger for assistance with the experiments and discussions. Part of this work is funded by the Deutsche Forschungsgemeinschaft (DFG, German Research Foundation)—Project-ID 201269156-SFB 1032 (Project B01 and B12).

# Supplementary Material for

## Inferring three-body interactions in cell migration dynamics

Agathe Jouneau,[1] Tom Brandstätter,[2,3] Bram Hoogland,[2] Joachim O. Rädler,[1] and Chase P. Broedersz[2]

[1]*Faculty of Physics and Center for NanoScience, Ludwig-Maximilians-Universität München, 80539 Munich, Germany*
[2]*Department of Physics and Astronomy, Vrije Universiteit Amsterdam, 1081 HV Amsterdam, The Netherlands*
[3]*Arnold-Sommerfeld-Center for Theoretical Physics, Ludwig-Maximilians-Universität München, 80333 Munich, Germany*

# EXPERIMENTAL METHODS

*Micropatterning*—The patterns are printed using a subtractive photopatterning technique. Ibidi µ-Dishes ibiTreat (ibidi GmbH, ref. 81156) are plasma-treated with $O_2$ for 1 min, and then incubated with a 100 µg/mL PLL solution (Sigma-Aldrich, ref. SLCP1100, diluted in water) for 30 min. The dishes are then washed with 0.1 M HEPES (pH 8.3–8.5) and incubated for one hour with 100 mg/mL mPEG-SVA (Laysan Bio, MPEG-SVA, MW 5,000) dissolved in HEPES (pH 8.3–8.5). The dishes are profusely washed with water and left to air dry. For each dish, 3 µL of the photoactivatable reagent PLPP Gel (Alvéole) is diluted in 60 µL of 99.9% ethanol. The gel is spread onto the dish surface and allowed to dry in the dark. Patterns are printed with a PRIMO device (Alvéole) through a 20x objective on a Nikon Ti inverted microscope using the Leonardo software (Alvéole), with a dose of 15 mJ/mm². The dishes are profusely washed with water, and then incubated in PBS for 5 min. The dishes are then incubated for 15 min with 20 µg/mL of fibronectin (YO Proteins, 663 Fibronectin (human) 5 mg, resuspended in PBS), with a ratio 1:3 of fibronectin labeled with Alexa Fluor$^{TM}$ 647 NHS-Ester (Invitrogen, ref. A37573). The dishes are washed with PBS and stored in the fridge overnight.

The patterns consist of 35 µm × 35 µm squares linked by 40 µm × 7 µm bridges.

*Cell culture*—MCF10A cells (ATCC, ref. CRL-10317) are cultured at 37°C and 5% CO2 in DMEM/F-12 with GlutaMAX supplement (Gibco, ref. 10565-018), additionally supplemented with 5% horse serum (Sigma-Aldrich, ref. H1270), 10 µg/mL insulin (Sigma-Aldrich, ref. I9278), 500 ng/mL hydrocortisone (Sigma-Aldrich, ref. H6909), 100 ng/mL cholera toxin (Sigma-Aldrich, ref. C8052) and 20 ng/mL human epidermal growth factor (Sigma-Aldrich, ref. E9644).

MDA-MB-231 stably transfected with H2B mCherry (gift from Betz Lab, University of Göttingen, Germany) are cultured at 37°C in L-15 medium with GlutaMAX supplement (Gibco, ref. 31415-086) with 10% FBS (Gibco, ref. 10437028).

The cells are kept in T25 flasks (Sarstedt AG, ref. 83.3910.300) and passaged every two to three days. For passaging, the cells are washed with PBS and incubated with Accutase (Invitrogen, ref. 00-4555-56) until they all detach. Culture medium is added and the solution is centrifuged at 500 rcf for 6 min for MCF10A cells, or 800 rcf for 3 min for MDA-MB-231 cells. The cell pellet is resuspended in culture medium and approximately $5 \cdot 10^5$ cells are added to the new flask.

*Cell seeding on micropatterns*—For experiments, approximately 15,000 cells are added per µ-dish and left to adhere in the incubator in culture medium. Once the cells have adhered on the patterns (after up to 5 h), the medium is exchanged. For the MCF10A cells, the medium is replaced by culture medium with 25 nM Hoechst 33342 for nuclear staining. For the MDA-MB-231 H2B mCherry cells, the medium is replaced by L-15 without phenol red (Gibco, ref. 21083-027) supplemented with 10% FBS. During experiments, MCF10A cells are kept in a 5% CO2 atmosphere and both MCF10A and MDA-MB-231 H2B mCherry cells are kept at 37 °C.

*Image acquisition*—Time-lapses are acquired over 48 h with a Nikon Eclipse Ti inverted microscope equipped with a Perfect Focus System using a 10x objective. The samples are placed in a temperature and $CO_2$ control chamber (Okolab). Fields of view where three cells are adhered onto the same pattern are selected. A fluorescence image of the patterns (made visible by the fibronectin labeled with Alexa 647) is acquired prior to starting the time-lapse. During the time-lapse, phase contrast and fluorescence images of the cell nuclei (made visible by the Hoechst stain for MCF10A cells, and by the H2B mCherry for the MDA-MB-231 cells) are acquired for each field of view, every 10 min.

*Image analysis*— Patterns containing three adhered cells are manually selected in space and time using a tailor-made program. If two of the three cells originate from the division of one cell, the selection begins once the two daughter cells have visibly spread and adhered to the pattern. If one of the three cells undergoes division during the experiment, the selection is stopped 60 minutes before the cell rounds up. We exclude patterns where one or more cells exhibit visible abnormalities, such as multiple or disrupted nuclei.

For each selection, the pattern is automatically detected and the centers of the cell nuclei are tracked using a second tailor-made Python program. Pattern detection uses the OpenCV library, while cell tracking uses the Trackpy library [44]. The results of the tracking algorithm are examined one by one and any errors made by the program (e.g. two cells being exchanged during tracking) are manually corrected. A total of 182 (resp. 73) trajectories were collected for the MDA-MB-231 (resp. MCF10A) cell line, with a total length of 1669 (resp. 2185) hours.

*Control data*— The control data for the two cells interacting in a two-island pattern used in Fig. 2 were obtained using the same methods. These data were previously used in the publication [17].

# THEORETICAL METHODS

*Inference method*—We infer the terms of the second-order differential equation of the main text from the measured cell trajectories using the Underdamped Langevin Inference method extensively described in [33]. In practice, we use the package UnderdampedLangevinInference, which is available on GitHub [45]. We add new basis functions to the code, in order to infer three-body interactions.

Before performing the inference on the data of three interacting cells, we verify that we can capture the dynamics of single cells migrating on our new three-island pattern. This confirms that Equation (1) is suited to describe single cell dynamics on this new pattern, despite the fact that it does not include a memory kernel. Memory is needed to capture the dynamics of freely migrating cells, but is not required to capture the dynamics of a cell migrating in a small, confining environment [35].

For the system with three interacting cells, we first optimize the inference for a purely pairwise interaction model. The acceleration of each cell is then given by:

$$\frac{dv_i}{dt} = F_{env}(x_i, v_i) + \sum_{j \neq i} [f(|\Delta x_{ij}|)\Delta x_{ij} + \gamma(|\Delta x_{ij}|)\Delta v_{ij}] + \sigma \eta_i(t) \tag{S1}$$

As in [12], we infer the single-cell term $F_{env}(x_i, v_i)$ using a basis consisting of trigonometric functions of $x_i$ and polynomials of $v_i$:

$$F_{env}(x_i, v_i) \approx \sum_{n=1}^{N} \sum_{m=1}^{M} \left[ A_{nm} \cos\left(\frac{2\pi n x_i}{w}\right) + B_{nm} \sin\left(\frac{2\pi n x_i}{w}\right) \right] v_i^m \tag{S2}$$

and the pairwise interaction terms using a basis consisting of exponential decays:

$$f(|\Delta x_{ij}|) = \sum_{k=1}^{K} C_k \exp\left(-\frac{|\Delta x_{ij}|}{k r_0}\right), \quad \gamma(|\Delta x_{ij}|) = \sum_{k=1}^{K} D_k \exp\left(-\frac{|\Delta x_{ij}|}{k r_0}\right) \tag{S3}$$

We compare the statistics obtained for simulations after inference with the experimental statistics for different values of $N$, $M$, $w$, $K$ and $r_0$. We select a set of parameters for which the statistics are best captured: $N = M = K = 3$ (the same value as in [12]), $w = 225$ μm (reflecting the size of our pattern), and $r_{max} = K r_0 = 25$ μm. We keep these values for the inference with an additional three-body interaction.

*Three-body interaction ansatzes*—In principle, in our framework, a three-body interaction induced by cells $j$ and $k$ on cell $i$ could be represented by any function $F_{3B}(x_i, x_j, x_k, v_i, v_j, v_k)$. However, due to technical limitations of the inference method and to visualization issues, it is infeasible to infer a function with six dependencies. We therefore take the following constraining assumptions:

- The first constraint enforces mirror symmetry of the system and its invariance under a reversal of the *x*-axis. Reversing the *x*-axis gives: $F_{3B}(-x_i, -x_j, -x_k, -v_i, -v_j, -v_k) = -F_{3B}(x_i, x_j, x_k, v_i, v_j, v_k)$.
- The second constraint is that all cells are treated as identical. When acting on cell $i$, cells $j$ and $k$ play an identical role, so $F_{3B}(x_i, x_j, x_k, v_i, v_j, v_k) = F_{3B}(x_i, x_k, x_j, v_i, v_k, v_j)$.
- We assume translational symmetry and only consider three-body interactions that depend only on the relative positions and velocities of the cells. We can therefore rewrite $F_{3B}(x_i, x_j, x_k, v_i, v_j, v_k) = \widetilde{F_{3B}}(\Delta x_{ij}, \Delta x_{ik}, \Delta v_{ij}, \Delta v_{ik})$, where $\Delta x_{ij} = x_i - x_j$ is the distance between the cell nuclei, and $\Delta v_{ij} = v_i - v_j$ the difference in velocities. Note that $\Delta x_{jk} = \Delta x_{ik} - \Delta x_{ij}$ and therefore does not need to be explicitly included in the arguments of the function (the same applies to $\Delta v_{jk}$).
- We further assume that the interactions are local, such that they go to zero when either $\Delta x_{ij}$ or $\Delta x_{ik}$ goes to infinity.

We test different ansatzes in accordance with these assumptions. The following table lists all the ansatzes and parameters tested. In the table, $g$ is the function inferred from the experimental data by ULI. For the ansatzes A, B, C and D, $g$ is a function of one variable. For the ansatzes E and F, $g$ is a function of two variables.

| Ansatz | Functional form | Parameters | Interpretation |
|---|---|---|---|
| A | $F_{3B} = g\left(\frac{\Delta x_{12} + \Delta x_{13}}{2}\right) \mathcal{G}(\Delta x_{23})$ $= g\left(\frac{\Delta x_{12} + \Delta x_{13}}{2}\right) e^{-\frac{\Delta x_{23}^2}{2\sigma^2}}$ | $K = 4$, $r_{max} \in \{50 \text{ μm}, 75 \text{ μm}\}$, $\sigma \in \{20 \text{ μm}, 30 \text{ μm}, 40 \text{ μm}\}$. | The collective effect of cells 2 and 3 on cell 1 when cells 2 and 3 are close together. |
| B | $F_{3B} = g\left(\frac{\Delta x_{12} + \Delta x_{13}}{2}\right) \mathcal{G}(|\Delta x_{23}| - d)$ $= g\left(\frac{\Delta x_{12} + \Delta x_{13}}{2}\right) e^{-\frac{(|\Delta x_{23}| - d)^2}{2\sigma^2}}$ | $K = 4$, $r_{max} \in \{50 \text{ μm}, 75 \text{ μm}, 100 \text{ μm}\}$, $\sigma \in \{20 \text{ μm}, 30 \text{ μm}, 40 \text{ μm}\}$, $d \in \{75 \text{ μm}, 150 \text{ μm}\}$. | The collective effect of cells 2 and 3 on cell 1 when cells 2 and 3 are at a distance $d$ from one another. |
| C | $F_{3B} = g(\Delta x_{12}) \mathcal{G}(\Delta x_{13}) + g(\Delta x_{13}) \mathcal{G}(\Delta x_{12})$ $= g(\Delta x_{12}) e^{-\frac{\Delta x_{13}^2}{2\sigma^2}} + g(\Delta x_{13}) e^{-\frac{\Delta x_{12}^2}{2\sigma^2}}$ | $K = 4$, $r_{max} = 75$ μm, $\sigma \in \{15 \text{ μm}, 30 \text{ μm}\}$. | The effect of cell 3 on cell 1 when cells 1 and 2 are close together. |
| D | $F_{3B} = g(\Delta x_{12}) \mathcal{G}(\Delta x_{13} - d \cdot \text{sgn}(\Delta x_{12}))$ $+ g(\Delta x_{13}) \mathcal{G}(\Delta x_{12} - d \cdot \text{sgn}(\Delta x_{13}))$ | $K = 4$, $r_{max} = 75$ μm, $\sigma = 15$ μm, | The effect of cell 3 on cell 1 when cells 1 and 2 are at a distance $d$ from one another. |
| | $= g(\Delta x_{12}) e^{-\frac{(\Delta x_{13} - d \cdot \text{sgn}(\Delta x_{12}))^2}{2\sigma^2}}$ $+ g(\Delta x_{13}) e^{-\frac{(\Delta x_{12} - d \cdot \text{sgn}(\Delta x_{13}))^2}{2\sigma^2}}$ | $d \in$ $\{-15\ \mu m,\ 15\mu m,\ 75\ \mu m,\ 150\ \mu m\}$. | |
| E | $F_{3B} = g(\Delta x_{12}, \Delta x_{13})$ | $K \in \{3,4\}$, $r_{max} \in \{50\ \mu m,\ 75\ \mu m\}$, $\sigma = r_{max}/K$. | Three-body interaction on cell 1 depending on the relative position to cell 2 and cell 3. |
| F | $F_{3B} = g(\Delta x_{12}, \Delta x_{13})\Delta v_{12}$ $+ g(\Delta x_{13}, \Delta x_{12})\Delta v_{13}$ | $K \in \{3,4\}$, $r_{max} \in \{50\ \mu m,\ 75\ \mu m\}$, $\sigma = r_{max}/K$. | Three-body interaction on cell 1 that is linear in the relative velocities with respect to cells 2 and 3, and dependent on the relative distances to cells 2 and 3. |

$K$ defines the number of basis functions and $r_{max}$ their range. In the definition of ansatz D, sgn refers to the sign function.

For the figures of the main text, and Fig. S1, S2 and S9, we infer $g$ using a basis consisting of antisymmetric Gaussian kernels:

$$g(\Delta x) = \sum_{k=1}^{K} \alpha_k \left( \exp\left(-\frac{(\Delta x - k r_0)^2}{2 r_0^2}\right) - \exp\left(-\frac{(-\Delta x - k r_0)^2}{2 r_0^2}\right) \right)$$

where $\alpha_k$ are the coefficients inferred by ULI.

Fig. 2 and 3 of the main text correspond to the ansatz A with antisymmetric Gaussian kernels, $K = 4$, $r_{max} = K r_0 = 75\ \mu m$ and $\sigma = 30\ \mu m$. Fig. 4(a) correspond to ansatz B with antisymmetric Gaussian kernels, $K = 4$, $r_{max} = 100\ \mu m$, $d = 75\ \mu m$ and $\sigma = 25\ \mu m$; and Fig. 4(b) to ansatz C with antisymmetric Gaussian kernels, $K = 4$, $r_{max} = 75\ \mu m$ and $\sigma = 15\ \mu m$.

*Error bars and inference accuracy*— We estimate error bars for the terms inferred by ULI with bootstrapping, as proposed in [40]. To do so, we create 100 bootstrap samples of the experimental data by resampling with replacement, and we perform inference on each sample. Then, we compute the standard deviation of the inferred term at each point of the function's domain. We define the error bar as ±3 standard deviation of the inferred interaction.

To check the accuracy of the inference method, we use simulations of trajectories with randomly chosen three-body interactions. Random three-body interactions $F_{3B}$ are constructed by assigning a random coefficient to each function in the basis set according to a uniform distribution $\mathcal{U}([-a, a])$. The maximum value $a$ is roughly estimated from the coefficients obtained when inferring the three-body interaction from the MDA-MB-231 experimental data, using the unbiased estimator for the maximum of a uniform law (also known as the German tank problem). As the absolute value of the largest of the four inferred coefficients is 30, we use $a = 40\ (\approx 30 + 30/4)$. The simulations are performed using the functions $F_{env}$ and $F_{2B}$ and the noise amplitude $\sigma$ inferred from a pairwise interaction model for each cell line. We simulate cellular trajectories according to Eq. (1) and Eq. (2) of the main text. For each cell line, we simulate trajectories of the same length as those in our experimental data set. We repeat this process 100 times, each time using a new randomly generated three-body interaction. For the ansatz presented in Fig. 2 of the main text, we find that in at least 95% of cases, the original three-body interaction lies within the bootstrap error bar (see Fig. S1 and S2).

Fig. S9 shows the results of the inference of three-body interactions for the four ansatzes A, B, C and D with error bars, for selected parameters. For a wider range of parameters and for the ansatzes E and F, we do not systematically compute a bootstrap error bar due to significant time costs. Nevertheless, we run the inference and use the inferred term to run simulations, as described in the next section. This allows us to see if the inferred three-body interaction improves how well the experimental data is captured by the simulations. As we do not observe a significant improvement in data capture for any of the ansatzes, we do not judge it necessary to investigate their error bars more thoroughly.

*Simulations*—We simulate trajectories according to Eq. (1) and Eq. (2) of the main text, using the Verlet integration method with a time step of $\Delta t = 10$ s. We subsample the resulting trajectories to obtain the same sampling rate and trajectory length as the experimental data. We initiate the simulation using random cell positions, and allow for a two-hour pre-run before data collection. We exclude simulations in which one or more cells exit the boundaries of the pattern.

*Autocorrelations and dwell times*—For each individual cellular trajectory $x(t)$, the normalized auto-correlation of positions is defined as $\rho_{xx}(t, t') = \langle \frac{(x(t) - \bar{x})(x(t') - \bar{x})}{\sigma^2} \rangle$ where $\bar{x}$ is the average position of the trajectory and $\sigma$ its standard deviation. The autocorrelation $\rho_{xx}(t, t')$ is then averaged on all observed trajectories. The same is done for the velocities. Dwell time is the amount of time that a given cell spends on one island of the pattern before moving to an adjacent island. To compute dwell times, the boundary of each island is set in the middle of the bridge connecting it to the neighboring island. The initial and final islands on which the cell is located are also included in the computation. As the experimental and simulated trajectories have the same length, this approximation introduces the same bias in both cases, allowing for direct comparison.

*Distribution of relative positions*—At each point in time, the cells are labelled 1, 2 and 3, in such a way that $x_1 > x_2 > x_3$. The distances $\Delta x_{12} = x_1 - x_2$ and $\Delta x_{13} = x_1 - x_3$ are measured and their joint distribution is plotted. In Fig. 3 and Fig. S7, the distribution of the simulated data is plotted symmetrically for ease of comparison.

*Transition between configurations*—The positions of the cell nuclei are discretized into three states: left, middle and right islands of the pattern. The boundaries are set in the middle of the bridges between the islands. The combined discretized positions of the three cells define a configuration. Since each cell can occupy one of three states, there is a total of 27 possible configurations. We measure the probabilities of transitions between configurations by recording whether cells transition from one configuration to another at each time step. For Fig. S8, we ignore configurations in which all three cells are on the same island since they are rare. We also take the symmetry of the system into account to reduce the number of configurations to four main configurations illustrated in the figure.

*Scores*—To compare the empirical distributions of the relative cell positions in the experimental and simulated data, we use the energy distance [46]. We record all values $(\Delta x_{12}, \Delta x_{13})$ in both the experimental and simulated data sets. Let $\boldsymbol{e}_1, \ldots, \boldsymbol{e}_n$ be the values from the experiment and $\boldsymbol{s}_1, \ldots, \boldsymbol{s}_n$ the values from the simulation. We compute $d(\boldsymbol{e}, \boldsymbol{s}) = \langle \| \boldsymbol{e}_i - \boldsymbol{s}_j \| \rangle$, $d(\boldsymbol{s}, \boldsymbol{s}) = \langle \| \boldsymbol{s}_i - \boldsymbol{s}_j \| \rangle$, and $d(\boldsymbol{e}, \boldsymbol{e}) = \langle \| \boldsymbol{e}_i - \boldsymbol{e}_j \| \rangle$ where $\langle . \rangle$ denotes the arithmetic average over all pairs $(i, j)$ and $\| . \|$ the Euclidean distance. The energy distance between the experimental and simulated data is then defined as $D(\mathrm{E}, S) = (2d(\boldsymbol{e}, \boldsymbol{s}) - d(\boldsymbol{e}, \boldsymbol{e}) - d(\boldsymbol{s}, \boldsymbol{s}))^{1/2}$.

To compare the transition rate between configurations, we take the Frobenius distance between the non-normalized transition matrices of the experimental and simulated data.

To make the comparison easier, we normalize both the energy distance of relative positions and the distance between the transition matrices in the following way: For each distance $D$, we determine a reference distance $D_{ref}$ between the experimental data and simulations based only a single-cell term $F_{env}$ and the noise term, but no cell-cell interaction term. We define the score as $1 - D/D_{ref}$. A negative score indicates that the simulation performs worse than a simulation without any interaction, whereas a score of 1 indicates that the simulation matches the experimental data perfectly. The error bars are computed by creating bootstrap samples of the experimental data and of the simulated data.

We also calculate the distances between the distributions of the cellular positions and velocities. However, the variation in distance observed between different bootstrap samples is much greater than the difference observed between different simulations. We therefore chose not to include them in the figures.

*Comparison of the scores for different three-body interactions*—For the MDA-MB-231 cell line, we carry out inference and simulations for the ansatzes A to F defined above. with the parameters given in the table.

For ansatzes A to D, we use basis functions consisting of antisymmetric Gaussian kernels:

$$g(\Delta x) = \sum_{k=1}^{K} \alpha_k (\exp(-\frac{(\Delta x - kr_0)^2}{2r_0^2}) - \exp(-\frac{(-\Delta x - kr_0)^2}{2r_0^2})),$$

and of exponential kernels:

$$g(\Delta x) = \sum_{k=1}^{K} \alpha_k \exp(-\frac{|\Delta x|}{kr_0}) \Delta x.$$

For the ansatz E, we use three different basis function expansions, chosen to respect the symmetries of the system:

$$g(\Delta x_{12}, \Delta x_{13}) = \sum_{k=0}^{K} \sum_{l=-k+1}^{k} \alpha_{k,l}(G_{kr_0, lr_0, \sigma}(\Delta x_{12}, \Delta x_{13}) + G_{lr_0, kr_0, \sigma}(\Delta x_{12}, \Delta x_{13}) + G_{-kr_0, -lr_0, \sigma}(\Delta x_{12}, \Delta x_{13}) + G_{-lr_0, -kr_0, \sigma}(\Delta x_{12}, \Delta x_{13})),$$

where $G_{x_0, y_0, \sigma}(x, y) = \exp\left(-\frac{(x-x_0)^2}{2\sigma^2} - \frac{(y-y_0)^2}{2\sigma^2}\right)$;

$$g(\Delta x_{12}, \Delta x_{13}) = \sum_{k=1}^{K} \sum_{l=1}^{K} \alpha_{k,l}(\exp\left(-\frac{|\Delta x_{12}|}{kr_0}\right) \cdot \exp\left(-\frac{|\Delta x_{13}|}{lr_0}\right) \Delta x_{12} + \exp\left(-\frac{|\Delta x_{12}|}{lr_0}\right) \cdot \exp\left(-\frac{|\Delta x_{13}|}{kr_0}\right) \Delta x_{13});$$

and

$$g(\Delta x_{12}, \Delta x_{13}) = \sum_{k=1}^{K} \sum_{l=1}^{K} \alpha_{k,l}(\exp\left(-\left(\frac{|\Delta x_{12}|}{kr_0}\right)^2\right) \cdot \exp\left(-\left(\frac{|\Delta x_{13}|}{lr_0}\right)^2\right) \Delta x_{12} + \exp\left(-\left(\frac{|\Delta x_{12}|}{lr_0}\right)^2\right) \cdot \exp\left(-\left(\frac{|\Delta x_{13}|}{kr_0}\right)^2\right) \Delta x_{13}).$$

Finally for the ansatz F, we use the three following expansions into basis functions:

$$g(\Delta x_{12}, \Delta x_{13}) = \sum_{k=0}^{K} \sum_{l=-k+1}^{k} \alpha_{k,l}(G_{kr_0, lr_0, \sigma}(\Delta x_{12}, \Delta x_{13}) + G_{kr_0, lr_0, \sigma}(-\Delta x_{12}, -\Delta x_{13}));$$

$$g(\Delta x_{12}, \Delta x_{13}) = \sum_{k=1}^{K} \sum_{l=1}^{K} \alpha_{k,l}(\exp\left(-\frac{|\Delta x_{12}|}{kr_0}\right) \cdot \exp\left(-\frac{|\Delta x_{13}|}{lr_0}\right));$$

and

$$g(\Delta x_{12}, \Delta x_{13}) = \sum_{k=1}^{K}\sum_{l=1}^{K} \alpha_{k,l}\left(\exp\left(-\left(\frac{|\Delta x_{12}|}{kr_0}\right)^2\right) \cdot \exp\left(-\left(\frac{|\Delta x_{13}|}{lr_0}\right)^2\right)\right).$$

We compare the scores of relative positions and transitions with the one of a simulation containing only pairwise interactions. For all the parameters tested, we find that simulations including three-body interactions do not perform significantly better than those based on a pairwise interaction model. The results are shown in Fig. S10.

## SUPPLEMENTARY MOVIES

**Supplementary Movie M1:** Several instances of three MCF10A cells interacting on a three-island micropattern. The cellular nuclei are labelled in blue to allow for automated tracking of cellular positions. The micropattern is labelled in red. Scale bar: 25 µm.

**Supplementary Movie M2:** Several instances of three MDA-MB-231 cells interacting on a three-island micropattern. The cellular nuclei are labelled in blue to allow for automated tracking of cellular positions. The micropattern is labelled in red. Scale bar: 25 µm.

**Supplementary Movie M3:** Simulation examples for three MCF10A cells interacting on a three-island micropattern. The green arrows represent the effective forces $F_i$ acting on each cell. Scale bars: green arrow: 100 µm/h²; white bar: 25 µm.

**Supplementary Movie M4:** Simulation examples for three MDA-MB-231 cells interacting on a three-island micropattern. The green arrows represent the effective forces $F_i$ acting on each cell. Scale bars: green arrow: 100 µm/h²; white bar: 25 µm.

**SUPPLEMENTARY FIGURES**

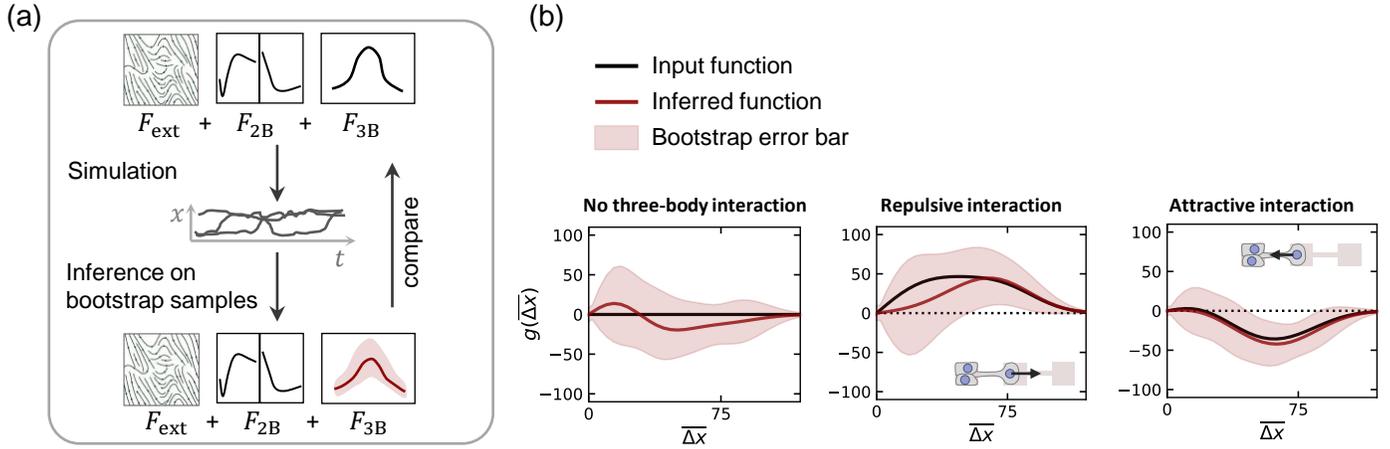

**Fig. S1. Validation of the three-body inference.** (a) To assess the validity and reliability of the three-body interaction inference, a randomly chosen three-body interaction $F_{3B}$ is used to simulate cell trajectories. The interaction is re-inferred from these trajectories with an error bar given by bootstrap sampling (in red). We verify that the original three-body interaction is encompassed in the error bar. (b) Inferred $F_{3B}$ and bootstrapping error bar for a simulation with zero three-body interaction and randomly chosen repulsive and attractive three-body interactions. For randomly chosen interactions, the input function lies within the error range in 95% of the cases.

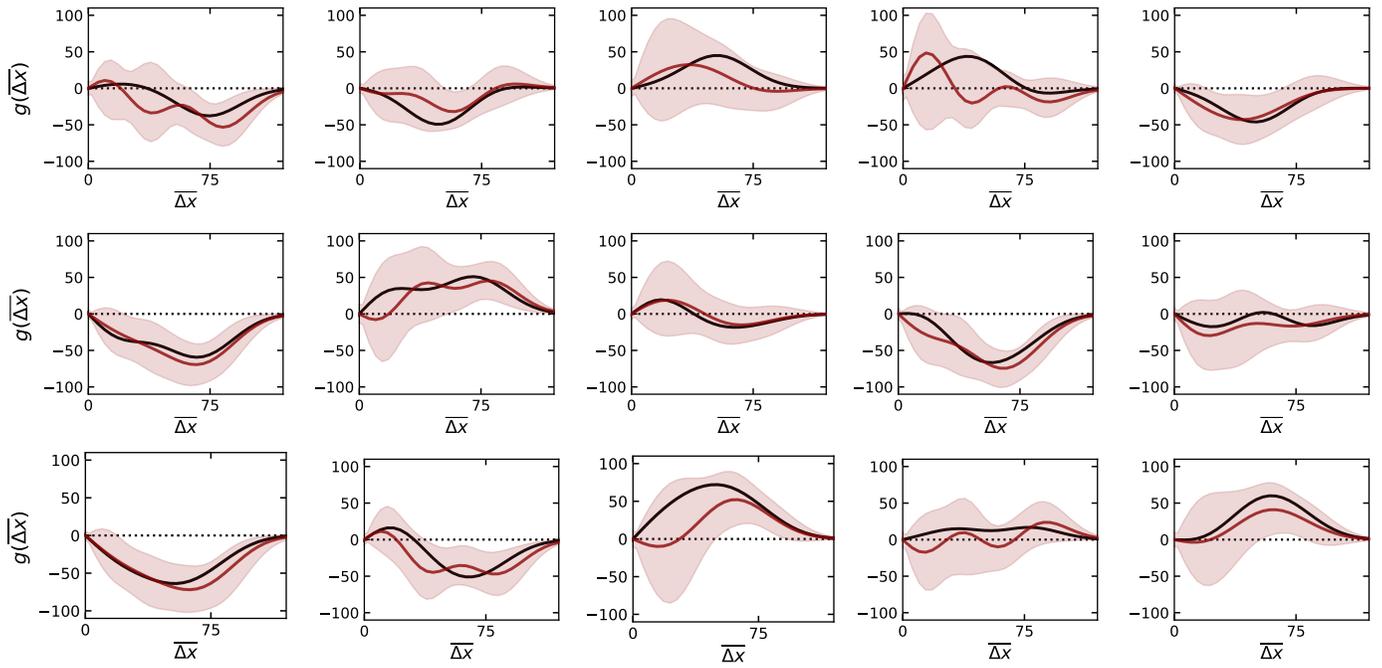

**Fig. S2. Validation of the three-body inference.** Additional examples of re-inference of three-body interactions from simulated data. The black lines show the randomly generated three-body interactions used for the simulation. The red lines show the three-body interactions re-inferred from the respective simulated trajectories. The light red zone shows the bootstrap error bar.

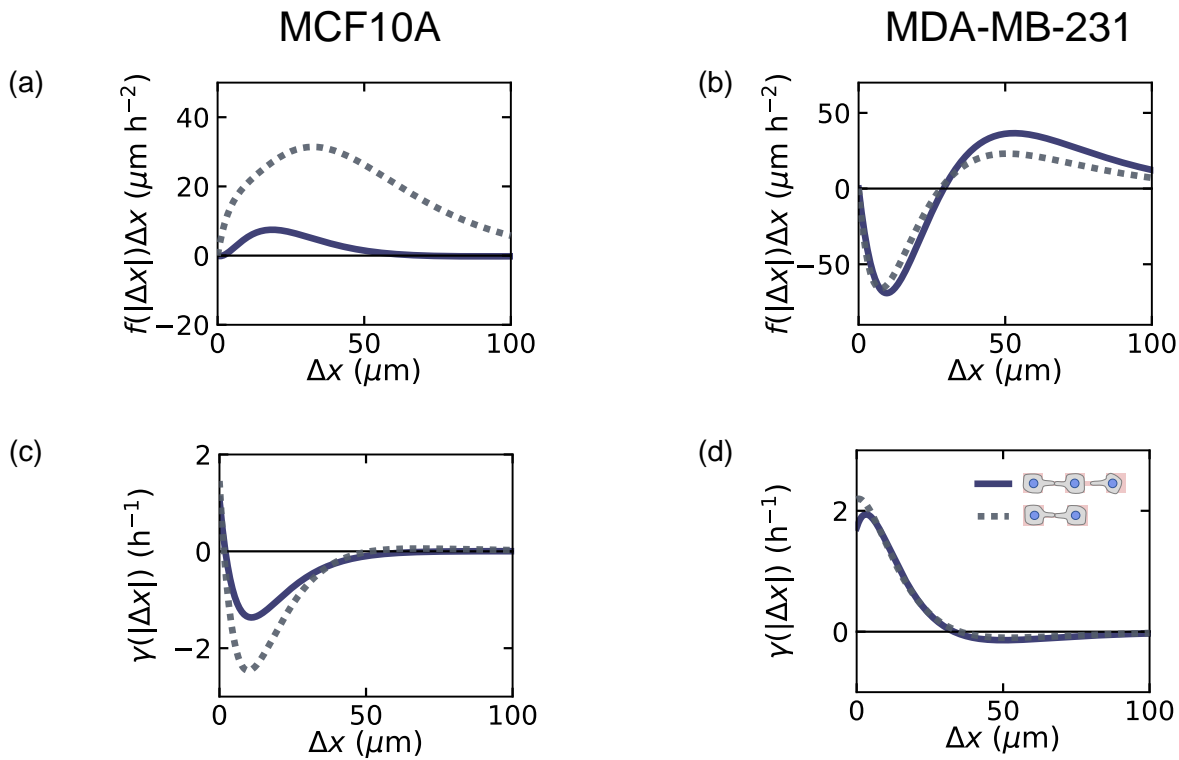

**Fig. S3. Comparison with pairwise interactions inferred in Brückner *et al.* [12]** (a and b) Cohesive interaction terms and (c and d) effective frictional interaction terms inferred for MCF10A and MDA-MB-231 cells. The solid blue lines are the terms that we infer for three interacting cells under the assumption of pairwise interactions. The dotted grey lines are the terms inferred in [12] for two cells interacting in a two-island confinement. The patterning method used in [12] differs from that used in the present study. In this figure, we use the parameter value $r_{max} = 20$ µm (see the *Inference method* section of the SM), as in [12].

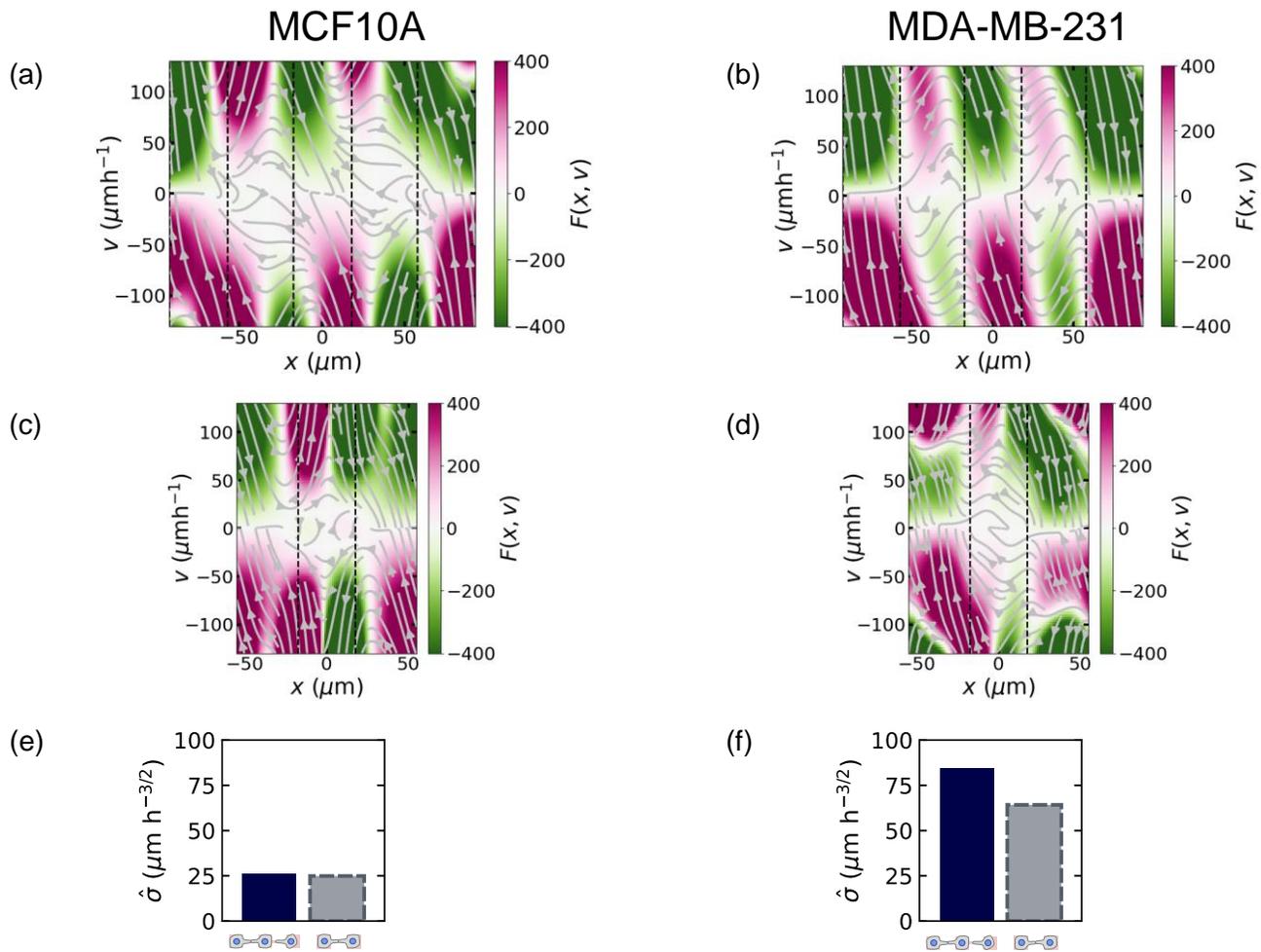

**Fig. S4. Comparison of single-cell term and noise amplitude inferred in three-cell system vs. two-cell system.** (a and b) Single-cell term inferred for three interacting cells for MCF10A and MDA-MB-231 cells, respectively. (c and d) Single-cell term inferred for two interacting cells in a two-island confinement for MCF10A and MDA-MB-231 cells. The black dashed lines show the boundary between the islands and the bridges. (e and f) Noise amplitudes inferred for three interacting cells in a three-island confinement (dark blue) and two interacting cells in a two-island confinement (gray with dashed border) for MCF10A and MDA-MB-231 cells.

(a) MCF10A
Experiment

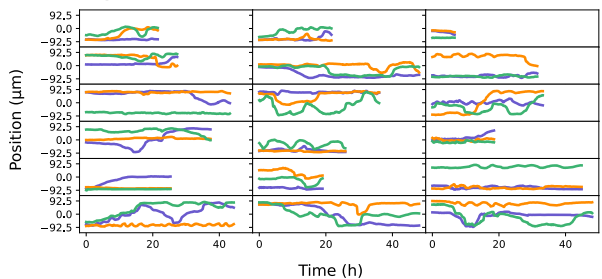

(b) MDA-MB-231
Experiment

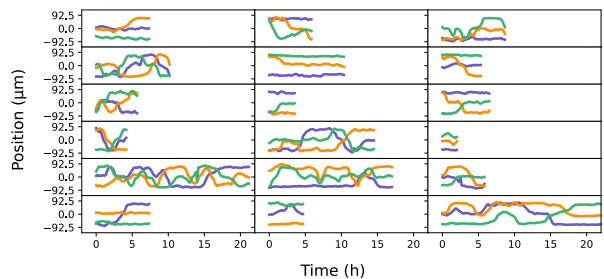

(c) MCF10A
Simulation
Pairwise interactions only

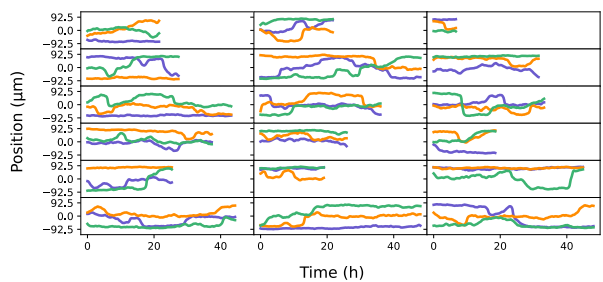

(d) MDA-MB-231
Simulation
Pairwise interactions only

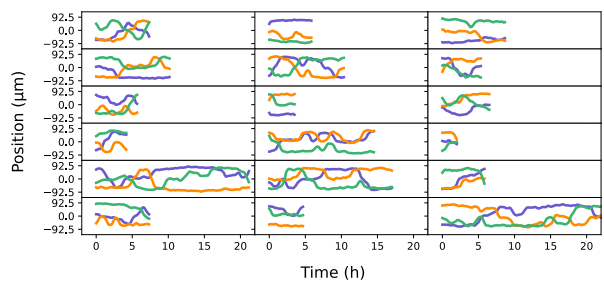

(e) MDA-MB-231
Simulation
Pairwise and three-body interactions

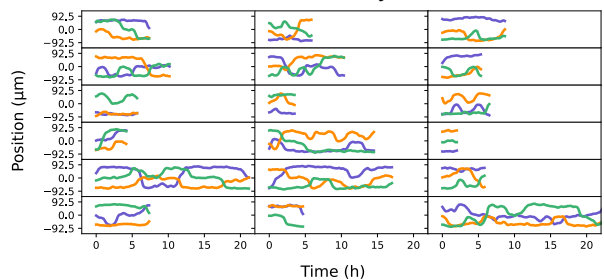

**Fig. S5. Experimental and simulated trajectories.** Subsets of trajectories for (a and b) experimental data and (c to e) simulated data presented in Fig. 3 of the main text.

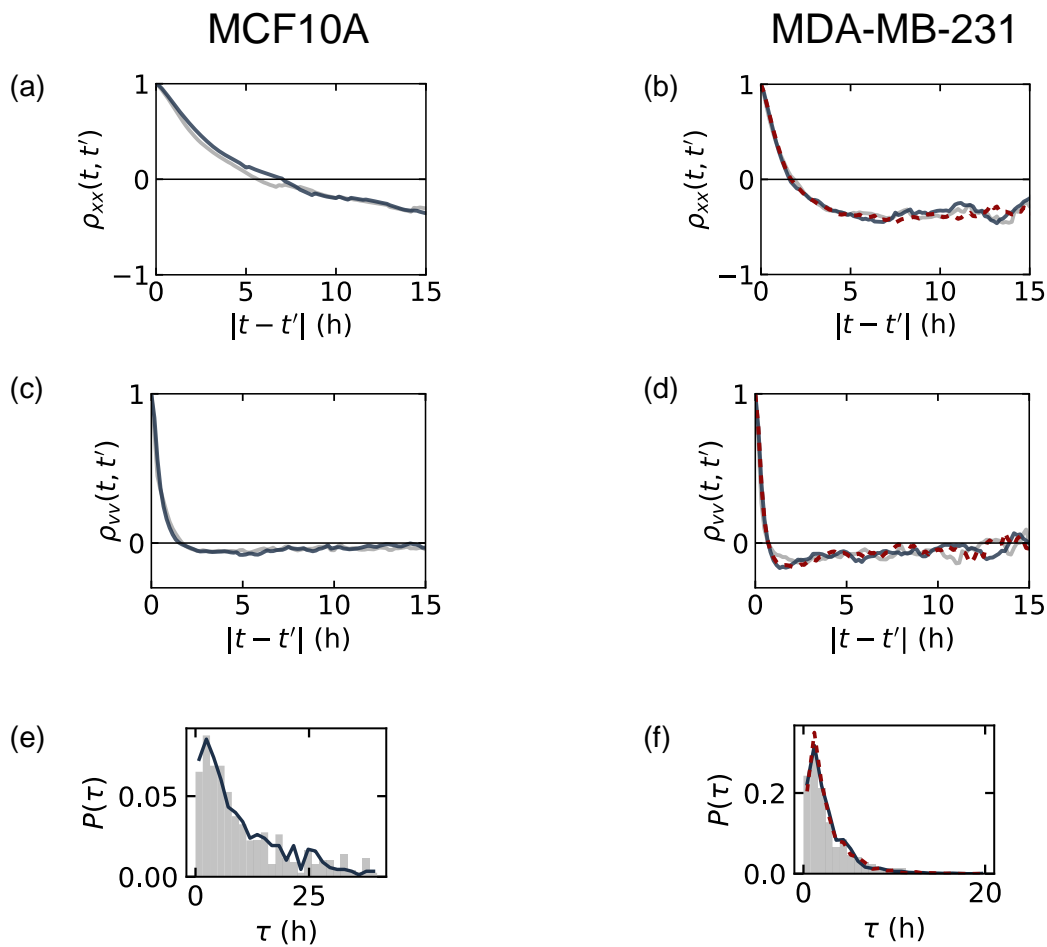

**Fig. S6. Comparison between simulated and experimental data.** (a and b) Position autocorrelation, (c and d) velocity autocorrelation and (e and f) distribution of dwell times $\tau$ (time spent by a cell on an island before moving to another island), for MCF10A and MDA-MB-231 cells. Grey lines and histograms show the experimental values. Continuous blue lines show the prediction of a pairwise interaction model and red dashed line of the model including the three-body interaction shown in Fig. 2(f) of the main text.

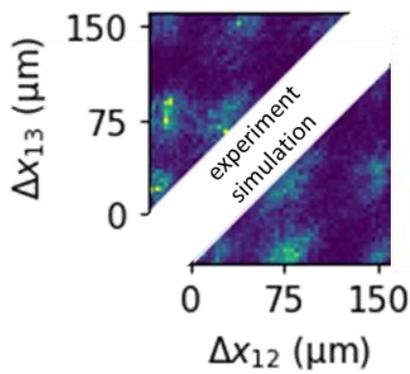 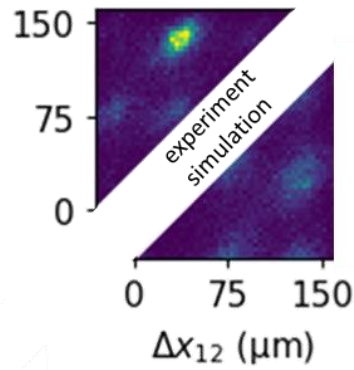 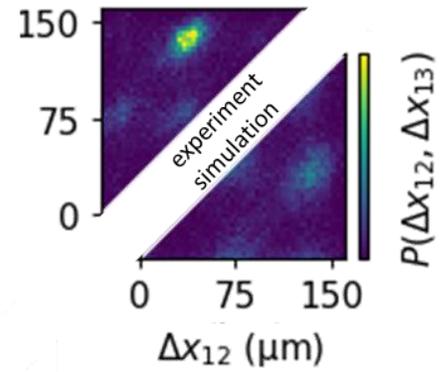

**Fig. S7 Joint distribution of the relative cell positions.** For each observed cellular configuration, we compute the distance between cell 1 and cell 2 $\Delta x_{12}$ and the distance between cell 1 and cell 3 $\Delta x_{13}$. We plot a two-dimensional histogram with the obtained values. The upper (respectively, lower) triangle shows the distribution in the experimental (respectively, simulated) data. (a): MCF10A cells, pairwise interactions model. (b): MDA-MB-231 cells, pairwise interactions model. (c): MDA-MB-231 cells, model including the three-body interaction shown in Fig. 2(f).

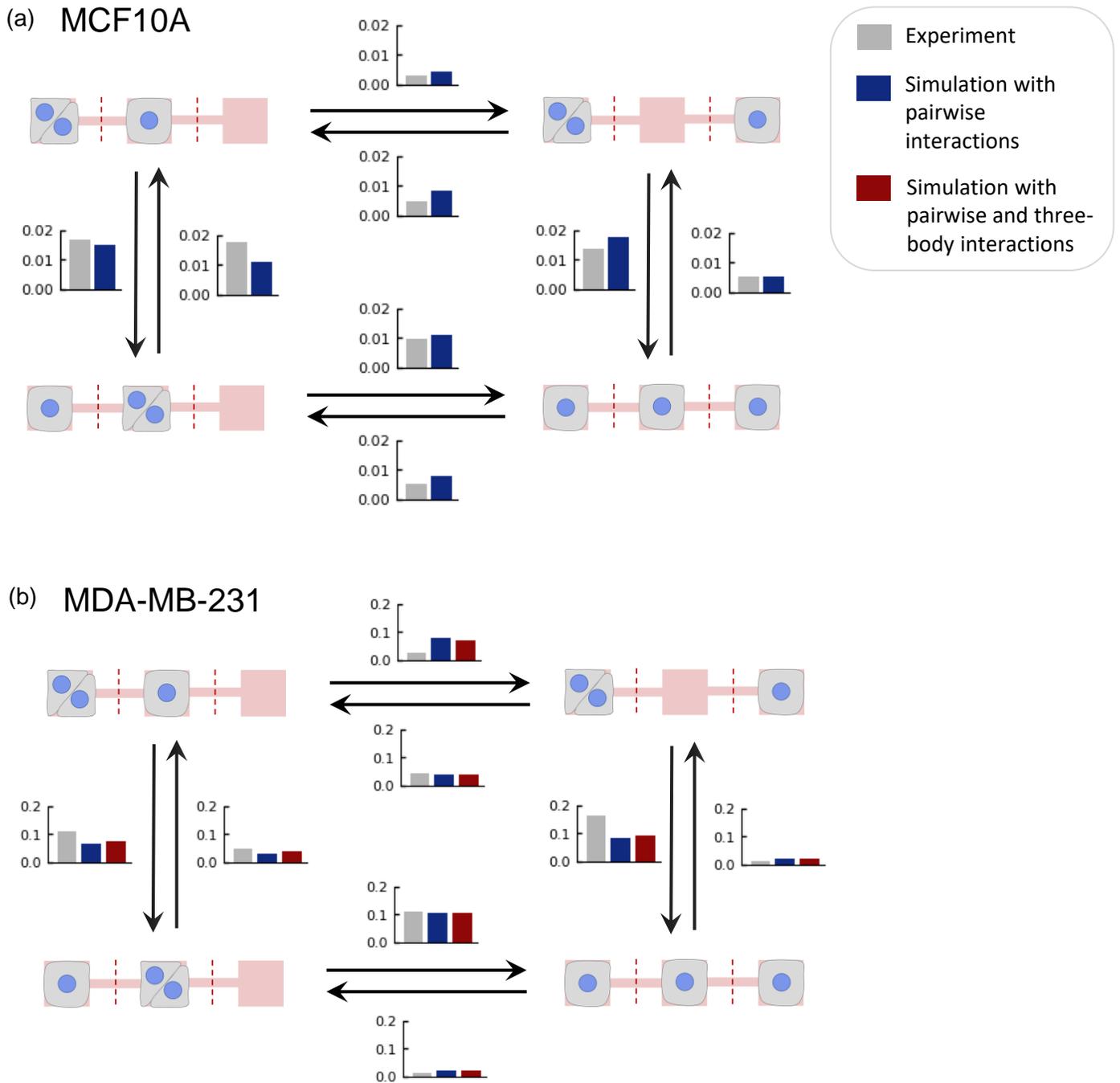

**Fig. S8. Transition rates between configurations for experimental and simulated data.** (a) and (b): The bar plots show the probability of transitioning between different configurations within the next 10 minutes for the cell lines MCF10A and MDA-MB-231, respectively. Configurations are defined by the position of the nuclear centers relative to the middle of the pattern bridges. Only the four most common configurations (illustrated in the sketches) and their associated transitions are shown. The grey bars show the probabilities for the experimental data; the dark blue bars show the probabilities for the simulation using a pairwise interaction model; and the red bars show the probabilities for the simulation that include the three-body interaction of Fig. 2(f).

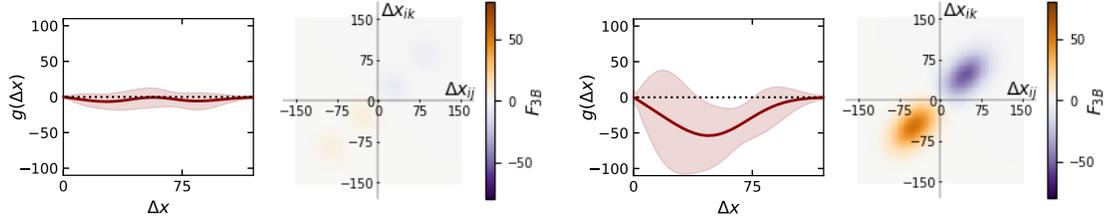
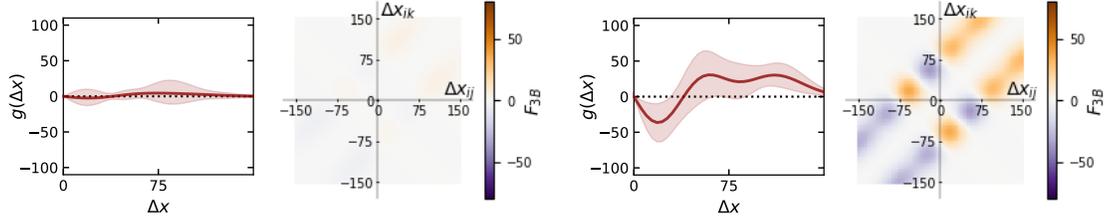
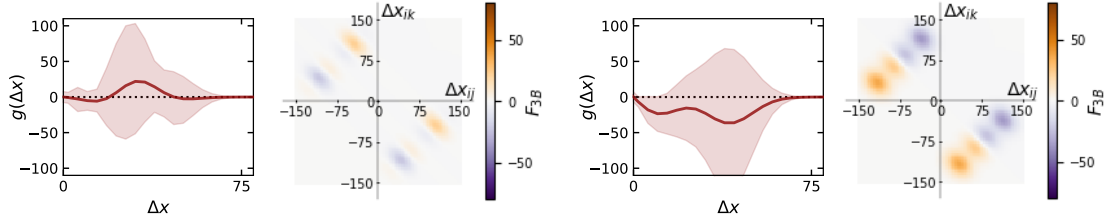
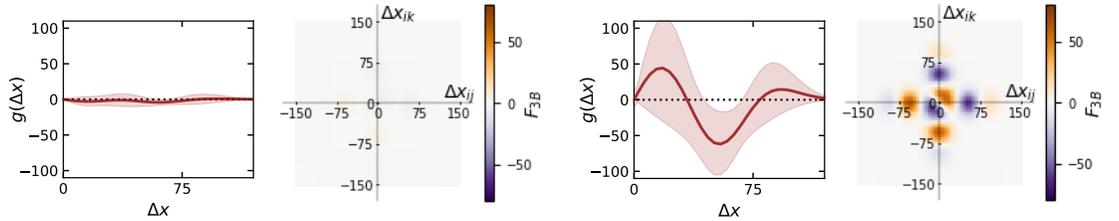
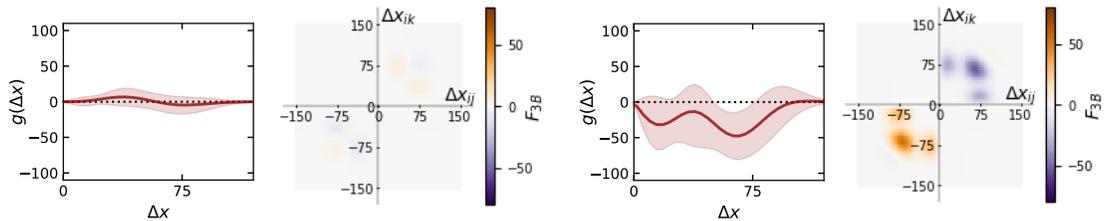
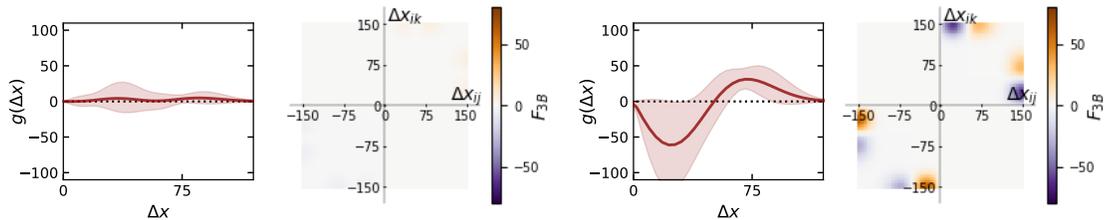
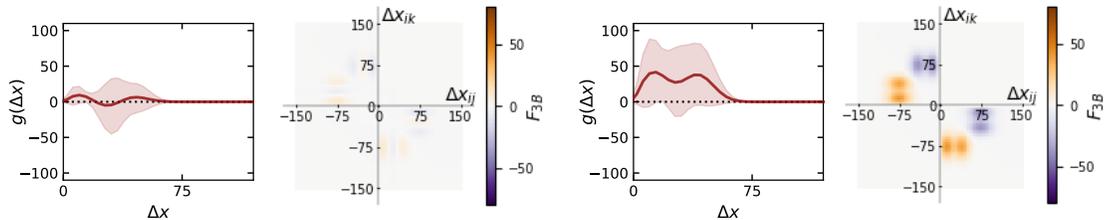

**Fig. S9. Inferred three-body interactions for different ansatzes.** The first column indicates which ansatz is used and the value of the parameters. The ansatzes A, B, C and D are formally defined in the section *Three-body interaction ansatzes* of the SM. For each ansatz and each cell line, the left-hand plot shows the inferred function $g(\Delta x)$, with the light red area corresponding to a 95% confidence interval. The right-hand plot is a two-dimensional representation of the function $F_{3B}(\Delta x_{12}, \Delta x_{13})$.

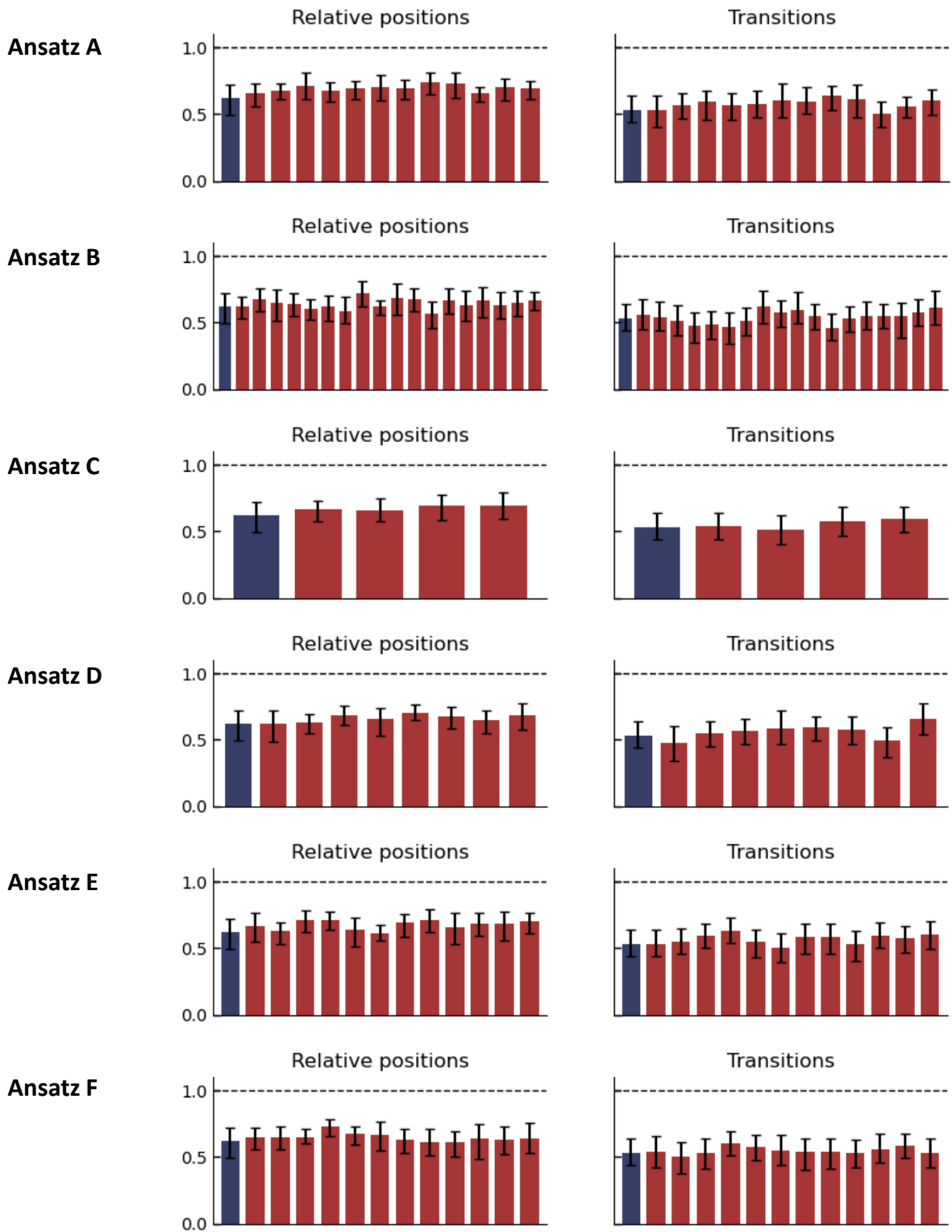

**Fig. S10. Score comparison for various three-body interaction ansatzes.** Scores obtained after performing three-body interaction inference and simulating trajectories, for the different ansatzes and parameters defined in the supplementary text. The first (resp. second) column are the scores of how well relative positions (resp. transitions between configurations) are captured (see definitions in the supplementary text) The red bars are the scores obtained for the simulations including three-body interactions, while the blue bars correspond to a simulation containing only pairwise interactions. The error bars represent a 95% interval calculated with bootstrapping.